\global\def\draftcontrol{0}

   \def\versionno{ jets -- draft   }
\catcode`\@=11

\expandafter\ifx\csname draftcontrol\endcsname\relax\global\def\draftcontrol{0}
\fi

{\count255=\time\divide\count255 by 60
\xdef\hourmin{\number\count255}
\multiply\count255 by-60\advance\count255 by\time
\xdef\hourmin{\hourmin:\ifnum\count255<10 0\fi\the\count255}}
\def\draftdate{\number\month/\number\day/\number\year\ \ \ \hourmin }

\newcommand\makepapertitle{\par
  \begingroup
    \renewcommand\thefootnote{\@fnsymbol\c@footnote}%
    \def\@makefnmark{\rlap{\@textsuperscript{\normalfont\@thefnmark}}}%
    \long\def\@makefntext##1{\parindent 1em\noindent
            \hb@xt@1.8em{%
                \hss\@textsuperscript{\normalfont\@thefnmark}}##1}%
     \newpage
     \global\@topnum\z@   % Prevents figures from going at top of page.
     \@makepapertitle
     \thispagestyle{empty}\@thanks
  \endgroup
  \setcounter{footnote}{0}%
  \global\let\thanks\relax
  \global\let\makepapertitle\relax
  \global\let\@makepapertitle\relax
  \global\let\@thanks\@empty
  \global\let\@author\@empty
  \global\let\@date\@empty
  \global\let\@title\@empty
  \global\let\title\relax
  \global\let\author\relax
  \global\let\date\relax
  \global\let\and\relax
  \def\version{\let\version\@version\@gobble}
}
\def\@makepapertitle{%
  \newpage
   \ifnum\draftcontrol=1 {}
   \version\versionno
   \vskip 3em%
   \else
   \hfill\hbox to 3cm {\parbox{4cm}{\@pubnum}\hss}%
   \vskip 3em%
   \fi
   \begin{center}%
   \let \footnote \thanks
     {\LARGE {\@title}}%
     \vskip 1.5em%
     {\normalsize%\large
       \lineskip .5em%
       \begin{tabular}[t]{c}%
         \@author
       \end{tabular}\par}%
     \vskip 1.5em%
     {\@bstract}%
     \end{center}%
     \vskip 1.5em
     \@date%
   \par
}

\gdef\@pubnum{}
\def\pubnum#1{%
  \gdef\@pubnum{#1}}

\gdef\@bstract{}
\def\Abstract#1{%
  \gdef\@bstract{%
   \parbox{\textwidth-0pc}{%
   \centerline{\bf Abstract}\penalty1000%
\kern.2cm%
\noindent%\abstractfont \baselineskip=12pt
\renewcommand\baselinestretch{1.0}%
{#1}}}
}

\def\ps@paper{\let\@mkboth\@gobbletwo%
     \ifnum\draftcontrol=1
    \def\@oddfoot{\hbox to \textwidth{\tiny \versionno \hfil\tiny\draftdate}%
    \hskip -\textwidth \hbox to \textwidth{\hfil\rm\thepage\hfil}}%
     \else\def\@oddfoot{\hbox to \textwidth{\hfil\rm\thepage\hfil}}
     \fi
     \let\@evenfoot\@oddfoot
}
\def\body{\clearpage
          \pagestyle{paper}
    }
\def\@version#1{\ifnum\draftcontrol=1
\typeout{}\typeout{#1}\typeout{}
\vskip3mm\centerline{\hbox{\fbox{\normalsize{\tt DRAFT -- #1 -- }
                   {\draftdate}}}}\vskip3mm
\fi}
\let\version\@version
\long\def\eqlabel#1{\ifnum\draftcontrol=1
                    \tag@false  % there are some problems with multline without this
                    \tag*{(\theequation) \hbox to -0.2cm{\hspace{0cm}\small{#1}\hss}}
                    \refstepcounter{equation}
                    \edef\@currentlabel{\theequation}
                    \ltx@label{#1}          % use old LaTeX \label instead of new definition
                    \else
                    \label{#1}
                    \fi
                    }
\let\st@bibitem\@bibitem
\let\st@lbibitem\@lbibitem
\ifnum\draftcontrol=1
  \def\@bibitem#1{%
    \st@bibitem{#1}\a@@label{#1}\ignorespaces}
  \def\@lbibitem[#1]#2{%
    \st@lbibitem[#1]{#2}\a@@label{#2}\ignorespaces}
  \def\a@@label#1{%
    \gdef\a@lab{\smash{\normalfont\small#1}}
    \ifvmode
      \if@inlabel
        \global\setbox\@labels\hbox{%
          \llap{\a@lab\let\a@lab\relax
                \kern\@totalleftmargin\kern\marginparsep}%
          \box\@labels}%
      \fi
    \fi}
\fi
\documentclass[12pt,letterpaper]{article}
\usepackage{amsmath,amssymb,array,calc,epsfig}
\ifnum\draftcontrol=1
\fi
\renewcommand\baselinestretch{1.25}
\renewcommand\section{\@startsection {section}{1}{\z@}%
                                   {-3.5ex \@plus -1ex \@minus -.2ex}%
                                   {2.3ex \@plus.2ex}%
                                   {\normalfont\large\bfseries}}
\renewcommand\subsection{\@startsection{subsection}{2}{\z@}%
                                   {-3.25ex\@plus -1ex \@minus -.2ex}%
                                   {1.5ex \@plus .2ex}%
                                   {\normalfont\normalsize\bfseries}}
\renewcommand\subsubsection{\@startsection{subsubsection}{3}{\z@}%
                                   {-3.25ex\@plus -1ex \@minus -.2ex}%
                                   {1.5ex \@plus .2ex}%
                                   {\normalfont\normalsize\it}}
\renewcommand\paragraph{\@startsection{paragraph}{4}{\z@}%
                                   {-3.25ex\@plus -1ex \@minus -.2ex}%
                                   {1.5ex \@plus .2ex}%
                                   {\normalfont\normalsize\bf}}

\numberwithin{equation}{section}

\def\revise#1       {\raisebox{-0em}{\rule{3pt}{1em}}%
                     \marginpar{\raisebox{.5em}{\vrule width3pt\
                     \vrule width0pt height 0pt depth0.5em
                     \hbox to 0cm{\hspace{0cm}{%
                     \parbox[t]{4em}{\raggedright\footnotesize{#1}}}\hss}}}}

\newcommand\nxt[1]  {\\\fnxt#1}

\def\calc         {{\cal C}}

\def\cali         {{\cal I}}

\def\calm         {{\cal M}}
\def\caln         {{\cal N}}
\def\calo         {{\cal O}}

\def\del          {\partial}

\def\sqr#1#2{{\vcenter{\vbox{\hrule height.#2pt
 \hbox{\vrule width.#2pt height#1pt \kern#1pt
 \vrule width.#2pt}\hrule height.#2pt}}}}

\newcommand{\ft}[2]{{\textstyle{\frac{#1}{#2}}}}

\def\a{\alpha}
\def\b{\beta}

\def\r{\rho}
\def\dd{\delta}

\def\ga{\gamma}

\def\hc{\hat{c}}
\def\hq{\hat{q}}

\def\aa1{\phi}
\def\cc1{\psi}

\def\G{\Gamma}

\def\l{\lambda}
\def\om{\Omega}
\def\dilog{\rm dilog}
\def\t{\tau}
\def\s{\sigma}
\catcode`\@=12

\begin{document}

%%%
%%%%%% text starts here
%%%%%%%%%

\title{On jet quenching parameters in strongly coupled non-conformal gauge theories}

\pubnum{%
UWO-TH-06/07}
\date{May 2006}

\author{
Alex Buchel\\[0.4cm]
\it Department of Applied Mathematics\\
\it University of Western Ontario\\
\it London, Ontario N6A 5B7, Canada\\[0.2cm]
\it Perimeter Institute for Theoretical Physics\\
\it Waterloo, Ontario N2J 2W9, Canada\\
}

\Abstract{Recently Liu, Rajagopal and Wiedemann (LRW) \cite{lrw} proposed a
first principle, nonperturbative quantum field theoretic definition of
``jet quenching parameter'' $\hq$ used in models of medium-induced
radiative parton energy loss in nucleus-nucleus collisions at
RHIC. Relating $\hq$ to a short-distance behavior of a certain
light-like Wilson loop, they used gauge theory-string theory
correspondence to evaluate $\hq$ for the strongly coupled $\caln=4$
$SU(N_c)$ gauge theory plasma. We generalize analysis of LRW to
strongly coupled non-conformal gauge theory plasma. We find that a jet
quenching parameter is gauge theory specific (not universal).
Furthermore, it appears it's value increases as the number of
effective adjoint degrees of freedom of a gauge theory plasma
increases.  
 }

\makepapertitle

\body

\version\versionno

\section{Introduction}

A characteristic feature of nuclear-nuclear collisions  actively studied at RHIC is QCD plasma-induced  ``jet quenching'' of partons 
produced with high transverse momentum \cite{rev1,rev2,rev3,rev4}. Successful models explaining such suppression of hadronic 
spectra typically involve a medium-sensitive ``jet quenching parameter'' $\hq$.  In\footnote{Alternative 
approaches to describing  jet quenching in hot gauge theory 
plasma where proposed in \cite{alt1,alt2}. Potential relation between two approaches 
has been explored in \cite{cg}.} \cite{lrw} Liu, Rajagopal and Wiedemann 
 proposed a first principle, nonperturbative quantum field theoretic definition of $\hq$. Specifically, they 
considered an expectation value of the light-like Wilson loop $\langle W^A(\calc)\rangle$ in the adjoint representation 
whose contour $\calc$ is a rectangle with large extension $L^-$ in the $x^-$ direction and small extension $L$ 
in a transverse direction. Motivated by so-called dipole approximation  used in jet quenching calculations \cite{dap}
\begin{equation}
 \langle W^A(\calc)\rangle\ \approx {\rm exp}\ \left[-\ft 14 \hq L^- L^2\right]\,,
\end{equation}
they  {\it defined} $\ft 14 \hq$ as the coefficient of the  $L^- L^2$ term in  $ \ln\langle W^A(\calc)\rangle $ at small $L$.
Quark-gluon plasma produced in nuclear-nuclear collisions at RHIC is believed to be strongly coupled. Given precise definition 
of $\hq$ as above, LRW further argued that gauge theory-string theory correspondence of Maldacena \cite{m9711,m2} is a suitable 
framework for such a computation. 
 
Currently, we do not have a useful string theoretic description of strongly coupled QCD ---  one typically studies ``QCD-like'' large-$N_c$ 
gauge theories that allow for a weakly curved supergravity description. By far the most popular model is $\caln=4$ $SU(N_c)$ supersymmetric 
Yang-Mills theory, which in the planar limit ( $N_c\to \infty$, $g_{YM}\to 0$ with $\l=g_{YM}^2 N_c={\rm const} $ ),
and for large values of the 't Hooft coupling  ( $\l \gg 1$ ), is described by type IIB supergravity on $AdS_5\times S^5$ \cite{m9711}.  
More refined examples are gauge-gravity dualities with less supersymmetries \cite{pw,ps,ks,mn}, or with fundamental matter \cite{kk,ss}.   
Though QCD itself is not in the universality class of these (and similar) models, one hopes that certain features 
of real QCD can be studied in this approximation. Relevant to study of strongly coupled gauge theory 
plasma, such hopes are supported by recent observation that all large-$N_c$ strongly coupled gauge theories
that have a dual supergravity description have a universal ratio  \cite{un1,un2,un3} of shear viscosity $\eta$ to the entropy 
density $s$,   
\begin{equation}
\frac {\eta}{s}=\frac{1}{4\pi}
\eqlabel{etas}
\end{equation}
in the limit\footnote{Leading  $\ft 1\l$ corrections to this ratio for $\caln=4$ $SU(N_c)$ SYM were 
computed in \cite{alp1,alp2}.} $\l\to \infty$. Also, though neither the speed of sound waves $v_s$ nor the bulk viscosity $\zeta$ 
in non-conformal gauge theories is universal, one observes an interesting phenomenological relation \cite{bul1,bul2,bul3}
\begin{equation}
\biggl(v_s^2-\frac 13\biggr)\ll 1\,,\qquad 
\frac{\zeta}{\eta}\simeq -\kappa\ \biggl(v_s^2-\frac 13\biggr)\,,\qquad \kappa\sim 1\,.
\eqlabel{nonconsf}
\end{equation}
An interesting question is how the jet quenching parameter $\hq$ fits into this story.
LRW found that for $\caln=4$ $SU(N_c)$ SYM at strong coupling and in the 't Hooft limit, the jet quenching parameter is
\begin{equation}
\hq_{\caln=4}= \frac{\pi^{3/2}\G\left(\ft 34\right)}{\sqrt{2}\G\left(\ft 54\right)}\ \sqrt{\l}\ T^3  \,,
\end{equation}
where $T$ is the temperature of the Yang-Mills plasma.  Furthermore, assuming that 
\begin{equation}
\hq_{QCD}\ \approx \hq_{\caln=4} \,,
\eqlabel{qqcd}
\end{equation}
and using a reasonable set of parameters relevant for Au-Au collisions at RHIC,  the jet quenching parameter 
predicted by \eqref{qqcd} appear to be somewhat small compare to the jet quenching parameter $\overline{\hq}$
extracted from the experiment \cite{lrw}. The authors suggested two possible explanations of the observed discrepancy:
\nxt QCD is not a conformal gauge theory; (one way) it is related to CFT is by reducing number of adjoint degrees
of freedom, thus LRW conjectured that in a process of such reduction a jet quenching parameter {\it increases}\ ;
\nxt  alternatively, they pointed out that $\overline{\hq}$ as extracted from the experiment could be misleadingly high because  
energy loss sources besides gluon radiation ( as the only source of energy loss assumed in evaluating $\hq$ ) could be important. 

In this paper, following  LRW proposal \cite{lrw},  
we study the jet quenching parameters in non-conformal gauge theories. We find that $\hq$ is not universal 
in strongly coupled gauge theory plasma in the 't Hooft limit that allow for an effective weakly curved supergravity description.
On the example of cascading gauge theory  \cite{ks}, we explicitly show that the jet quenching parameter increases
as one goes from a confining gauge theory to a conformal gauge theory. Thus, is appears  that a discrepancy between 
$\hq_{QCD}$ (in approximation \eqref{qqcd}) and $\overline{\hq}$ is likely due to additional energy loss sources 
for the hadronic jets besides gluon radiation.  

In the next section we recall relevant facts for the cascading gauge theory \cite{ks} and the 
supergravity dual of it's strongly coupled plasma \cite{kt1,kt2,kt3,kt4}. In section 3, following the general suggestion of \cite{lrw} 
we evaluate the jet quenching parameter of this cascading gauge theory $\hq_{cascade}$ and compare it to  $\hq_{\caln=4}$. 
We conclude in section 4 with  a proposal how a jet quenching parameter computed 
in the cascading gauge theory could be adapted to real QCD.   

Though our emphasis in this paper is on a cascading gauge theory, our analysis can be extended to other non-conformal gauge theory plasma, 
such as $\caln=2^*$ plasma \cite{n21,n22}. The jet quenching parameter in $\caln=2^*$ model will be discussed elsewhere.

\section{Cascading gauge theory}

In this section we recall the relevant facts about cascading gauge theories \cite{ks} and 
the gravitational description of their strongly coupled deconfined plasma \cite{kt1,kt2,kt3,kt4}.
In particular, we emphasize why cascading gauge theory plasma is an excellent 'probe' of the behavior 
of the jet quenching parameter $\hq$ as one goes from QCD to conformal plasma.    

\subsection{Gauge theory description}
Cascading gauge theory at a given high-energy scale resembles $\caln=1$ supersymmetric $SU(K_*)\times SU(K_*+P)$ 
gauge theory with two bifundamental and two anti-fundamental chiral superfields and certain superpotential,
which is quartic in superfields. 
Unlike ordinary quiver gauge theories, an 'effective rank' of  cascading gauge theories depends on an energy scale 
at which the theory is probed \cite{kt1,kt3,kt4}
\begin{equation}
K_*\equiv K_*(E)\sim 2 P^2\ \ln\frac{E}{\Lambda}\,,\qquad E\gg \Lambda\,.
\eqlabel{kdep}
\end{equation}
At a given temperature $T$ cascading gauge theory is probed at energy scale $E\sim T$, 
and as $T\gg \Lambda$, $K_*(T)\gg P^2$. In this regime the thermal properties of the theory \cite{kt3,kt4} are very similar to 
those of the $\caln=1$ $SU(K_*)\times SU(K_*)$ superconformal gauge theory of Klebanov and Witten \cite{kw},
with 
\begin{equation}
\dd_{cascade}\equiv \frac{P^2}{K_*} 
\eqlabel{defcas}  
\end{equation}
being the deformation parameter. As the temperature increases, the deformation parameter $\dd_{cascade}$ decreases 
and the theory more and more resembles conformal gauge theory. The latter is probably 
best illustrated by the behavior of the cascading gauge theory plasma transport properties, such as the speed of sound 
and the bulk viscosity \cite{bul2}
\begin{equation}
\begin{split}
&v_s^2=\frac 13-\frac 49\ \dd_{cascade}+\calo(\dd_{cascade}^2)\,,\\
&\frac{\zeta}{\eta} = - 2 \, \left(
v_s^2-\frac{1}{3}\right)+\calo\biggl(\dd_{cascade}^2\biggr)\,.
\end{split}
\eqlabel{ratiobulk}
\end{equation}
At small temperatures (below the strong coupling scale)
the cascading gauge theory, much like real QCD,  is expected to confine and undergo chiral symmetry breaking\footnote{This was rigorously 
demonstrated only at zero temperature \cite{ks}. }. Thus, the temperature dependence of the ratio
\begin{equation}
\r(T) = \frac{\hq_{cascade}}{\hq_{KW}}\,,
\eqlabel{r}
\end{equation} 
as one dials up the temperature is a good indicator of  what happens with the jet quenching parameter as one goes from 
QCD to conformal Klebanov-Witten plasma with the jet quenching parameter $\hq_{KW}$. 

In section 3 we explicitly evaluate $\hq_{cascade}$ at high temperature and show that 
\begin{equation}
\frac{d\r(T)}{dT} >0\,, \qquad T\gg \Lambda \,.
\eqlabel{dr}
\end{equation}

\subsection{Supergravity dual to strongly coupled deconfined cascading gauge theory plasma}
Supergravity dual to deconfined cascading gauge theory plasma was studied in \cite{kt1,kt2,kt3,kt4}.
The ten-dimensional Einstein frame metric takes a form 
\begin{equation}
ds_{10(E)}^2 =g_{\mu\nu} dy^{\mu}dy^{\nu}+\om_1^2(r) e_{\psi}^2 
+\om_2^2(r) \sum_{i=1}^2\left(e_{\theta_i}^2+e_{\phi_i}^2\right),
\eqlabel{10met}
\end{equation}
where $r$ is the radial  coordinate on $\calm_5$ (greek indexes $\mu,\nu$
will run from $0$ to $4$) and
the one-forms $e_{\psi},\ e_{\theta_i},\ e_{\phi_i}$ ($i=1,2$) are given by :
\begin{equation}
e_{\psi}=\frac 13 \left(d\psi+\sum_{i=1}^2 \cos\theta_i\ 
d\phi_i\right),\qquad
e_{\theta_i}=\frac{1}{\sqrt{6}} d\theta_i\,,\qquad 
e_{\phi_i}=\frac{1}{\sqrt{6}} \sin\theta_i\ d\phi_i\,.
\eqlabel{1forms}
\end{equation} 
The five-dimensional metric can be conveniently parameterized as 
\begin{equation}
g_{\mu\nu} dy^{\mu}dy^{\nu}=\om_1^{-2/3}\om_2^{-8/3}\ \biggl(-c_1^2\ dt^2+ c_2^2\ d\vec{x}^2+c_3^2\ dr^2\biggr)\,,
\eqlabel{met5}
\end{equation}
with $c_i=c_i(r)$.
The metric \eqref{met5}  has a regular horizon at $c_1(r)=0$.
Finally, the background geometry also has nontrivial 5-form and 3-form fluxes and the dilaton $\Phi(r)$.
Introducing a new radial coordinate 
\begin{equation}
x\equiv \frac{c_1}{c_2}\,,
\eqlabel{radgauge}
\end{equation}
(the horizon is now at $x=0$ and the boundary is at $x=1$)
and 
\begin{equation}
\om_1=e^{f-4w}\,,\qquad \om_2=e^{f+w}\,,
\eqlabel{defom12}
\end{equation}
explicit geometry and the dilaton to leading order in $\dd_{cascade}$ are determined by
\begin{equation}
\begin{split}
&c_2=\frac{a}{(1-x^2)^{1/4}}\ \left(1+\frac{P^2}{K_*}\xi(x)\right)\,,\qquad f=-\frac 14\ln\frac{4}{K_*}
+\frac{P^2}{K_*}\eta(x)\,,\\
&w=\frac{P^2}{K_*} \psi(x)\,,\qquad \Phi=\frac{P^2}{K_*}\zeta(x)\,,
\end{split}
\eqlabel{solvebackP2}
\end{equation}
where $a$ is a constant nonextremality parameter, and 
\begin{equation}
\begin{split}
&\xi=\frac{1}{12}(1-\ln(1-x^2))\,,\\
&\zeta=\frac{K_*}{P^2}{\Phi_{horizon}} +\frac{\pi^2}{12}-\frac 12\dilog(x)+\frac 12 \dilog(1+x)
-\frac 12 \ln x\ln (1-x)\,,\\
&\eta=-\frac{3(1+x^2)}{80(1-x^2)}\left(\dilog(1-x^2)-\frac {\pi^2}{6}\right)+\frac{1}{20}(1-\ln(1-x^2))\,.
\end{split}
\eqlabel{expP2}
\end{equation}
Furthermore, $\psi$ satisfies the linear differential equation 
\begin{equation}
0=\psi''+\frac 1x \psi'-\frac{3}{(1-x^2)^2}\psi-\frac{1}{10(1-x^2)}\,,
\eqlabel{eqom}
\end{equation}
with the boundary condition
\begin{equation}
\psi=\psi_{horizon}+\calo(x^2)\,,\qquad \psi=-\frac{1}{30}(1-x^2)+\calo\biggl((1-x^2)^{3/2}\biggr)\,,
\eqlabel{asspsi}
\end{equation}
where the second boundary condition will uniquely determine $\psi_{horizon}$.
In what follows we fix the dilaton so that 
\begin{equation}
\lim_{x\to 1_- }\ \Phi=0\,,
\end{equation}
which leads to
\begin{equation}
\Phi_{horizon}=-\frac{\pi^2P^2}{24 K_{*}}\,.
\eqlabel{dhor}
\end{equation}

The  Hawking temperature $T$ of the nonextremal solution (again to leading order in $\dd_{cascade}$) is given by 
\begin{equation}
(2\pi T)^2= 2^{14/3}\ a^2\ K_{*}^{-4/3}\left(1-\frac{P^2}{2K_{*}}\right)\,.
\eqlabel{tempeature}
\end{equation}

\section{Calculation of $\hq_{cascade}$}
Following \cite{lrw}, the jet quenching parameter is determined from the expectation value of a certain light-like Wilson loop 
in the adjoint representation  $ \langle W^A(\calc)\rangle $ . On the supergravity side it is easiest to evaluate 
the thermal   expectation value of   a  Wilson loop in the fundamental representation $ \langle W^F(\calc)\rangle $. 
 In the planar limit the two expectation values are related as follows
\begin{equation}
 \ln\langle W^A(\calc)\rangle = 2\ \ln\langle W^F(\calc)\rangle\,.
\end{equation}
According to the gauge theory-string theory correspondence \cite{wl1,wl2,wl3,wl4,wl5},  $\langle W^F(\calc)\rangle$ 
is given by 
\begin{equation}
\langle W^F(\calc)\rangle={\rm exp}\ \left[-S(\calc)\right]\,,
\eqlabel{defwf}
\end{equation}
where $S$ is the extremal action (subject to a suitable subtraction)
 of a fundamental string worldsheet whose $r\to \infty$ boundary is the $\calc$ contour in Minkowski space $R^{3,1}$.

Using light-cone Minkowski coordinates $(t,\vec{x})=(x^\pm,x^2,x^3)$, the relevant contour $\calc$ is then a rectangle with 
large extension $L^-$ in the $x^-$ direction and small extension $L$ along the $x^2$ direction. 
As in \cite{lrw}, we parameterized the surface whose action $S(\calc)$ is to be extremized by 
\[
\{\xi^I(\t,\s)\}=\biggl\{x^{\pm}(\t,\s)\,,\ x^2(\t,\s)\,,\ x^3(\t,\s)\,,\ r(\t,\s)\,,\ \psi(\t,\s)\,,\ \theta_i(\t,\s)\,,\ \phi_i(\t,\s)\biggr\}\,,
\]
for $i=1,2$, where $\s^\a=(\t,\s)$ describe the coordinates parameterizing the worldsheet. 
The Nambu-Goto action for the string worldsheet is given by 
\begin{equation}
S=\frac{1}{2\pi\a'}\int d\s d\t \sqrt{{\rm det}g_{\a\b}}\,,
\eqlabel{sng}
\end{equation}
with
\begin{equation}
g_{\a\b}=G_{IJ}\ \del_\a \xi^I\ \del_\b \xi^J\,,
\eqlabel{inducedmet}
\end{equation}
where $G_{IJ}$ is the string frame metric of the background ten-dimensional geometry. 
In our case 
\begin{equation}
\begin{split}
&ds_{10(string)}^2=e^{\Phi/2}\ ds_{10(E)}^2= G_{IJ}d\xi^Id\xi^J\\
=&-\left(\hc_1^2+\hc_2^2\right) dx^+dx^-+\frac 12\left(\hc_2^2-\hc_1^2\right)\left((dx^+)^2+(dx^-)^2\right)
+\hc_2^2\left(dx_2^2+dx_3^2\right)+\hc_3^2 dr^2\\
&+\hat{\om}_1^2 e_{\psi}^2+\hat{\om}_2^2 \sum_{i=1}^2\left(e_{\theta_i}^2+e_{\phi_i}^2\right)\,,
\end{split}
\eqlabel{10dmetric}
\end{equation}
where 
\begin{equation}
\begin{split}
\hc_i^2=e^{\Phi/2}\ \om_1^{-2/3}\om_2^{-8/3}\  c_i^2\,,\qquad \hat{\om}_j^2=e^{\Phi/2}\ \om_j^2\,,
\end{split}
\eqlabel{hatdef}
\end{equation}
and $i=1,2,3$, $j=1,2$.
We will fix string worldsheet  coordinates as $\t=x^-$  and $\s=x^2$. In the limit $ L^- \gg L$
it is consistent to assume that (apart from $x^-$) $\xi^I=\xi^I(\s)$.  The symmetries of the background 
geometry imply that the extremal string worldsheet would lie at constant $x^+,x^3,\psi,\theta_i,\phi_i$. 
For the remaining bulk coordinate $r$, we implement the requirement that the world sheet has $\calc$ 
as its boundary by imposing 
\begin{equation}
r\left(\pm \frac L2\right)=\infty\,.
\eqlabel{rbound}
\end{equation} 
Notice that such an embedding preserves a symmetry $r(\s)=r(-\s)$. 
The action \eqref{sng} takes form
\begin{equation}
S=\frac{\sqrt{2}L^-}{2\pi\a'}\ \int_0^{\ft L2}d\s\ \hc_2^2\ \left(1-\frac{\hc_1^2}{\hc_2^2}\right)^{1/2}\ \left(1+\frac{\hc_3^2(r')^2}
{\hc_2^2}\right)^{1/2}\,,
\eqlabel{fo1}
\end{equation}
where $r'=\del_\s\ r$. The equation of motion for $r(\s)$ is then
\begin{equation}
\frac{\hc_3^2(r')^2}{\hc_2^2}=\ga\ \hc_2^4\ \left(1-\frac{\hc_1^2}{\hc_2^2}\right)-1\,,
\eqlabel{reom}
\end{equation}
where $\ga> 0$ is an  integration constant. This integration constant is determined by $L$, and as we will see shortly, 
small values of $L$ correspond to large values of $\ga$.  
The $\s\leftrightarrow -\s$ symmetry of the string worldsheet implies that $r'(\s=0)=0$. 
At the horizon  $r=r_h$ of the geometry \eqref{10dmetric} $\hc_1$ vanishes  and the left hand side of \eqref{reom}
is manifestly positive\footnote{At least for sufficiently large values of $\ga$.}. In fact, one can explicitly verify that  
for large enough $\ga$ it stays positive  all the way from the horizon to the boundary. Thus we conclude that at $\s=0$
the string worldsheet must reach the horizon (where $\hc_3(r_h)=\infty$)
\begin{equation}
r(\s=0)=r_h  \,.
\eqlabel{r0}
\end{equation}
Along with \eqref{reom}, the latter boundary condition relates $\ga$ to the transverse width $L$ of the Wilson loop
\begin{equation}
\frac L2=\int_{r_h}^{\infty}\ \frac{\hc_3\ dr}{\hc_2\ (\ga\ \hc_2^2\ (\hc_2^2-\hc_1^2)-1)^{1/2}}\,.
\eqlabel{lgarel}
\end{equation}
In the case of supergravity dual to $\caln=4$ plasma 
\[
\hc_2^2\ (\hc_2^2-\hc_1^2)-1
\]
is a constant\footnote{Given conformal invariance of the background geometry 
this is not surprising.}. In non-conformal gauge theories it varies; moreover, the combination multiplying 
$\ga$ never vanishes\footnote{This is obviously true for any asymptotically AdS or Klebanov-Tseytlin backgrounds.}. 
Thus we conclude from \eqref{lgarel} that large values of $L$ correspond to small values of $\ga$ (in non-conformal geometries). 

Using equation of motion \eqref{reom} we can rewrite 
\begin{equation}
S=\frac{\sqrt{2}L^-}{2\pi\a'}\ \int_{r_h}^{\infty}\ \frac{\ga^{1/2}\ \hc_2^2\ (\hc_2^2-\hc_1^2)\ \hc_3\ dr}
{\hc_2\ (\ga\ \hc_2^2\ (\hc_2^2-\hc_1^2)-1)^{1/2}}\,.
\eqlabel{fo2}
\end{equation}
As in \cite{lrw} from \eqref{fo2} one needs to subtract the self-energy $S_0$ of the high energy quark and antiquark moving 
through the plasma:
\begin{equation}
S_0=\frac{\sqrt{2}L^-}{2\pi\a'}\ \int_{r_h}^{\infty}\ dr\sqrt{G_{--}G_{rr}}=\frac{\sqrt{2}L^-}{2\pi\a'}\ \int_{r_h}^{\infty}\ 
\hc_3\ (\hc_2^2-\hc_1^2)^{1/2}\ dr\,.
\end{equation}
The resulting $S_I=S-S_0$ is the subtracted extremal action to be used in \eqref{defwf}.
Recall that we are interested in the expectation value of the thermal Wilson loop as $L\to 0$. We argued above that in that
limit $\ga\to\infty$, thus, rather explicitly, to leading order in $\ga$ we have
\begin{equation}
S_I=\frac{\sqrt{2}L^-}{2\pi\a'}\ \frac{1}{2\ga} \int_{r_h}^{\infty} \frac{\hc_3\ dr}{\hc_2^2\ (\hc_2^2-\hc_1^2)^{1/2}}\,,\qquad 
\frac L2=\frac{1}{\ga^{1/2}}\ \int_{r_h}^{\infty} \frac{\hc_3\ dr}{\hc_2^2\ (\hc_2^2-\hc_1^2)^{1/2}}\,.
\eqlabel{largeg}
\end{equation}
Notice that $S_I\propto L^2$, and the universality (or not) of the thermal Wilson loop expectation value 
(ans thus the jet quenching parameter) is related to the properties of the background integral $\cali$
\begin{equation}
\cali=\int_{r_h}^{\infty} \frac{\hc_3\ dr}{\hc_2^2\ (\hc_2^2-\hc_1^2)^{1/2}}\,.
\eqlabel{defi}
\end{equation}
Further simplification is possible by noticing that reduction from the ten dimensional string frame \eqref{10dmetric} 
to the five-dimensional Einstein frame\footnote{In our case the latter is simply\
$(-c_1^2\ dt^2+c_2^2\ \vec{x}^2+ c_3^2\ dr^2)$.} leads to 
\begin{equation}
\cali=\int_{r_h}^{\infty} e^{-\Phi/2}\ \om_1^{2/3}\ \om_2^{8/3}\  \frac{c_3\ dr}{c_2^2\ (c_2^2-c_1^2)^{1/2}}\,.
\eqlabel{defi1}
\end{equation}
The utility of the five dimensional Einstein frame is because in all gauge theory-string theory  dualities (subject to condition 
of \cite{un1}) one has 
\begin{equation}
\left[\ \frac{c_2^4}{c_3}\ \left[\frac{c_1}{c_2}\right]'\ \right]'=0\,,
\eqlabel{rel}
\end{equation}
where derivatives are with respect to $r$. 
Of course, equation \eqref{rel} can be verified explicitly in each example of duality --- for the supergravity dual to 
cascading gauge theory it follows from eq.(32) and eq.(33) of \cite{bul2}.
Integrating \eqref{rel} we find
\begin{equation}
\left[\frac{c_1}{c_2}\right]'=2\pi T\ c_{2,h}^3\ \frac{c_3}{c_2^4}\,.  
\eqlabel{xxx}
\end{equation}
where $c_{2,h}=c_2(r_h)$. Now, \eqref{xxx} allows to change the integration variable in \eqref{defi1} from $r$ to $x$ given 
by \eqref{radgauge}, we find
\begin{equation}
\cali=\frac{1}{2\pi T\ c_{2,h}^3}\ \int_0^1 dx\ \frac{c_2(x)}{\sqrt{1-x^2}}\ e^{-\Phi(x)/2}\ \om_1^{2/3}(x)\ \om_2^{8/3}(x) \,.
\eqlabel{defi3} 
\end{equation}
In a class of  gauge theory-string theory  dualities discussed in  
 \cite{un1} (and \cite{un3}) the part of the integrand in \eqref{defi3} depending on  the five-dimensional  metric
warp factor $c_2$ and the dilaton $\Phi$ is universal; the remaining factor is $\propto g_{\perp}^{1/3}$ , where 
$g_{\perp}$ is the determinant of the transverse (angular) Einstein frame metric\footnote{It is easy to reproduce $\caln=4$ result of 
\cite{lrw} from  our general approach.}.  Obviously, $\cali$ is not 
``universal''.  

We conclude this section with explicit evaluation of the jet quenching parameter in cascading gauge theories. 
Using background metric \eqref{solvebackP2} and  \eqref{expP2} we find (to leading order in $\dd_{cascade}$)
\begin{equation}
\begin{split}
\r(T)=\frac{\hq_{cascade}}{\hq_{KW}}=1+\frac{P^2}{K_*}\ \chi+\calo\left(\frac{P^4}{K_*^2}\right)\,,
\end{split} 
\eqlabel{ratio}
\end{equation}
with\footnote{In terms of supergravity parameters $\hq_{KW}$ is identical to that of $\hq_{\caln=4}$ computed in \cite{lrw}. However, 
these supergravity parameters have a different relation to gauge theory parameters in Klebanov-Witten and $\caln=4$ superconformal 
gauge theories. It can be shown  that $\frac{\hq_{KW}}{\hq_{\caln=4}}=\sqrt{\frac{27}{32}}$, due to the difference in the $S^5$ and $T^{1,1}$ 
volumes and the fact that Klebanov-Witten superconformal gauge theory contains two gauge groups \cite{hl}.} 
\begin{equation}
\chi=\frac 12+3\ \xi(0)-\frac{\sqrt{2}\left(\Gamma\left(\frac 34\right)\right)^2}{6\pi^{3/2}}\ \int_0^{1} dx\ 
\frac{6\ \xi(x)+20\ \eta(x)-3\ \zeta(x)}{(1-x^2)^{3/4}}\,.
\eqlabel{chival}
\end{equation}
Numerically evaluating the integral in \eqref{chival} we find
\begin{equation}
\chi\ \approx -1.388\,.
\eqlabel{chires}
\end{equation}
Since $\chi<0$  and $K_*$ increases with temperature \eqref{kdep}, we obtain \eqref{dr}.

\section{Jet quenching in QCD}
 \footnote{I would like to thank Krishna Rajagopal for insightful comments that led to this discussion.}
As we stated in the introduction, asymptotically free gauge theories are not in the universality class of gauge theories dual
to weakly coupled supergravity backgrounds. Nonetheless, it is interesting to see what would be the prediction of 
QCD jet quenching parameter $\hq_{QCD}$ from supergravity. We propose that such a relation could be obtained 
by expressing the temperature-dependent deviation of $\hq_{cascade}$ from $\hq_{\caln=4}$ in terms of the temperature-dependent
deviation of the speed of sound from the conformal result.  Combining \eqref{ratio} and \eqref{ratiobulk} we find
\begin{equation}
\hq_{cascade}=\hq_{KW}\ \times\ \biggl(1+\frac{9\chi}{4}\left(\frac 13-v_s^2\right)\ \biggr)
\eqlabel{jetqcd}
\end{equation}
Now, \eqref{jetqcd} could be adapted to real QCD replacing $\hq_{cascade}\Rightarrow \hq_{QCD}$, 
$\hq_{KW}\Rightarrow \hq_{\caln=4}$ and 
\[
v_s^2\ \Rightarrow \left(v_s^{QCD}\right)^2=\frac{\del P_{QCD}}{\del \epsilon_{QCD}}
\]
where $P_{QCD}(T)$ and  $\epsilon_{QCD}(T)$ are correspondingly the pressure and the energy density of QCD plasma 
in the regime relevant at RHIC. 

We expect that  $\chi$ computed from the cascading gauge theory would differ from the corresponding coefficient
extracted from  $\caln=2^*$ gauge theory plasma. It is thus interesting to obtain numerical value for $\hq_{QCD}$
from different dual supergravity models and see whether the result is robust.

\section*{Acknowledgments}
I would like to thank Ofer Aharony, Hong Liu and Krishna Rajagopal for valuable comments.
Research at Perimeter Institute is supported in part by funds from NSERC of
Canada. I gratefully   acknowledge  support by  NSERC Discovery
grant.


\begin{thebibliography}{99}

\bibitem{lrw}
  H.~Liu, K.~Rajagopal and U.~A.~Wiedemann,
  ``Calculating the Jet Quenching Parameter from AdS/CFT,''
  arXiv:hep-ph/0605178.
  %%CITATION = HEP-PH 0605178;%%

\bibitem{alt1} 
  J.~Casalderrey-Solana and D.~Teaney,
  ``Heavy quark diffusion in strongly coupled N = 4 Yang Mills,''
  arXiv:hep-ph/0605199.
  %%CITATION = HEP-PH 0605199;%%


\bibitem{alt2}
  C.~P.~Herzog, A.~Karch, P.~Kovtun, C.~Kozcaz and L.~G.~Yaffe,
   ``Energy loss of a heavy quark moving through N = 4 supersymmetric Yang-Mills
  plasma,''
  arXiv:hep-th/0605158.
  %%CITATION = HEP-TH 0605158;%%

\bibitem{cg}
  E.~Caceres and A.~Guijosa,
  ``On drag forces and jet quenching in strongly coupled plasmas,''
  arXiv:hep-th/0606134.
  %%CITATION = HEP-TH 0606134;%%


\bibitem{rev1}
  R.~Baier, D.~Schiff and B.~G.~Zakharov,
  ``Energy loss in perturbative QCD,''
  Ann.\ Rev.\ Nucl.\ Part.\ Sci.\  {\bf 50}, 37 (2000)
  [arXiv:hep-ph/0002198].
  %%CITATION = HEP-PH 0002198;%%

\bibitem{rev2}
  A.~Kovner and U.~A.~Wiedemann,
  ``Gluon radiation and parton energy loss,''
  arXiv:hep-ph/0304151.
  %%CITATION = HEP-PH 0304151;%%

\bibitem{rev3}
  M.~Gyulassy, I.~Vitev, X.~N.~Wang and B.~W.~Zhang,
  ``Jet quenching and radiative energy loss in dense nuclear matter,''
  arXiv:nucl-th/0302077.
  %%CITATION = NUCL-TH 0302077;%%

\bibitem{rev4}
  P.~Jacobs and X.~N.~Wang,
  ``Matter in extremis: Ultrarelativistic nuclear collisions at RHIC,''
  Prog.\ Part.\ Nucl.\ Phys.\  {\bf 54}, 443 (2005)
  [arXiv:hep-ph/0405125].
  %%CITATION = HEP-PH 0405125;%%

\bibitem{dap}
  B.~G.~Zakharov,
  ``Radiative energy loss of high energy quarks in finite-size nuclear  matter
  and quark-gluon plasma,''
  JETP Lett.\  {\bf 65}, 615 (1997)
  [arXiv:hep-ph/9704255].
  %%CITATION = HEP-PH 9704255;%%

\bibitem{m9711}J.~M.~Maldacena,
``The large $N$ limit of superconformal field theories and supergravity,''
Adv.\ Theor.\ Math.\ Phys.\  {\bf 2}, 231 (1998)
[Int.\ J.\ Theor.\ Phys.\  {\bf 38}, 1113 (1999)]
[arXiv:hep-th/9711200].


\bibitem{m2}
  O.~Aharony, S.~S.~Gubser, J.~M.~Maldacena, H.~Ooguri and Y.~Oz,
  ``Large N field theories, string theory and gravity,''
  Phys.\ Rept.\  {\bf 323}, 183 (2000)
  [arXiv:hep-th/9905111].
  %%CITATION = HEP-TH 9905111;%%

\bibitem{pw}
  K.~Pilch and N.~P.~Warner,
  ``N = 2 supersymmetric RG flows and the IIB dilaton,''
  Nucl.\ Phys.\ B {\bf 594}, 209 (2001)
  [arXiv:hep-th/0004063].
  %%CITATION = HEP-TH 0004063;%%


\bibitem{ps}
  J.~Polchinski and M.~J.~Strassler,
  ``The string dual of a confining four-dimensional gauge theory,''
  arXiv:hep-th/0003136.
  %%CITATION = HEP-TH 0003136;%%


\bibitem{ks}
  I.~R.~Klebanov and M.~J.~Strassler,
  ``Supergravity and a confining gauge theory: Duality cascades and
  chiSB-resolution of naked singularities,''
  JHEP {\bf 0008}, 052 (2000)
  [arXiv:hep-th/0007191].
  %%CITATION = HEP-TH 0007191;%%

\bibitem{mn}
  J.~M.~Maldacena and C.~Nunez,
  ``Towards the large N limit of pure N = 1 super Yang Mills,''
  Phys.\ Rev.\ Lett.\  {\bf 86}, 588 (2001)
  [arXiv:hep-th/0008001].
  %%CITATION = HEP-TH 0008001;%%

\bibitem{kk}
  A.~Karch and E.~Katz,
  ``Adding flavor to AdS/CFT,''
  JHEP {\bf 0206}, 043 (2002)
  [arXiv:hep-th/0205236].
  %%CITATION = HEP-TH 0205236;%%

\bibitem{ss} 
T.~Sakai and S.~Sugimoto,
  ``Low energy hadron physics in holographic QCD,''
  Prog.\ Theor.\ Phys.\  {\bf 113}, 843 (2005)
  [arXiv:hep-th/0412141].
  %%CITATION = HEP-TH 0412141;%%

\bibitem{un1}
  A.~Buchel and J.~T.~Liu,
  ``Universality of the shear viscosity in supergravity,''
  Phys.\ Rev.\ Lett.\  {\bf 93}, 090602 (2004)
  [arXiv:hep-th/0311175].
  %%CITATION = HEP-TH 0311175;%%

\bibitem{un2}
  P.~Kovtun, D.~T.~Son and A.~O.~Starinets,
  ``Viscosity in strongly interacting quantum field theories from black hole
  physics,''
  Phys.\ Rev.\ Lett.\  {\bf 94}, 111601 (2005)
  [arXiv:hep-th/0405231].
  %%CITATION = HEP-TH 0405231;%%

\bibitem{un3}
  A.~Buchel,
  ``On universality of stress-energy tensor correlation functions in
  supergravity,''
  Phys.\ Lett.\ B {\bf 609}, 392 (2005)
  [arXiv:hep-th/0408095].
  %%CITATION = HEP-TH 0408095;%%

\bibitem{alp1}
  A.~Buchel, J.~T.~Liu and A.~O.~Starinets,
  ``Coupling constant dependence of the shear viscosity in N = 4
  supersymmetric Yang-Mills theory,''
  Nucl.\ Phys.\ B {\bf 707}, 56 (2005)
  [arXiv:hep-th/0406264].
  %%CITATION = HEP-TH 0406264;%%

\bibitem{alp2}
  P.~Benincasa and A.~Buchel,
  ``Transport properties of N = 4 supersymmetric Yang-Mills theory at finite
  coupling,''
  JHEP {\bf 0601}, 103 (2006)
  [arXiv:hep-th/0510041].
  %%CITATION = HEP-TH 0510041;%%

\bibitem{bul1}
  P.~Benincasa, A.~Buchel and A.~O.~Starinets,
  ``Sound waves in strongly coupled non-conformal gauge theory plasma,''
  Nucl.\ Phys.\ B {\bf 733}, 160 (2006)
  [arXiv:hep-th/0507026].
  %%CITATION = HEP-TH 0507026;%%

\bibitem{bul2}
  A.~Buchel,
  ``Transport properties of cascading gauge theories,''
  Phys.\ Rev.\ D {\bf 72}, 106002 (2005)
  [arXiv:hep-th/0509083].
  %%CITATION = HEP-TH 0509083;%%

\bibitem{bul3}
  P.~Benincasa and A.~Buchel,
  ``Hydrodynamics of Sakai-Sugimoto model in the quenched approximation,''
  arXiv:hep-th/0605076.
  %%CITATION = HEP-TH 0605076;%%

\bibitem{kt1}
  A.~Buchel,
  ``Finite temperature resolution of the Klebanov-Tseytlin singularity,''
  Nucl.\ Phys.\ B {\bf 600}, 219 (2001)
  [arXiv:hep-th/0011146].
  %%CITATION = HEP-TH 0011146;%%

\bibitem{kt2}
  A.~Buchel, C.~P.~Herzog, I.~R.~Klebanov, L.~A.~Pando Zayas and A.~A.~Tseytlin,
  ``Non-extremal gravity duals for fractional D3-branes on the conifold,''
  JHEP {\bf 0104}, 033 (2001)
  [arXiv:hep-th/0102105].
  %%CITATION = HEP-TH 0102105;%%

\bibitem{kt3}
  S.~S.~Gubser, C.~P.~Herzog, I.~R.~Klebanov and A.~A.~Tseytlin,
  ``Restoration of chiral symmetry: A supergravity perspective,''
  JHEP {\bf 0105}, 028 (2001)
  [arXiv:hep-th/0102172].
  %%CITATION = HEP-TH 0102172;%%

\bibitem{kt4}
  O.~Aharony, A.~Buchel and A.~Yarom,
  ``Holographic renormalization of cascading gauge theories,''
  Phys.\ Rev.\ D {\bf 72}, 066003 (2005)
  [arXiv:hep-th/0506002].
  %%CITATION = HEP-TH 0506002;%%

\bibitem{n21}
  A.~Buchel and J.~T.~Liu,
  ``Thermodynamics of the N = 2* flow,''
  JHEP {\bf 0311}, 031 (2003)
  [arXiv:hep-th/0305064].
  %%CITATION = HEP-TH 0305064;%%

\bibitem{n22}
  A.~Buchel,
  ``N = 2* hydrodynamics,''
  Nucl.\ Phys.\ B {\bf 708}, 451 (2005)
  [arXiv:hep-th/0406200].
  %%CITATION = HEP-TH 0406200;%%

\bibitem{kw}
  I.~R.~Klebanov and E.~Witten,
  ``Superconformal field theory on threebranes at a Calabi-Yau  singularity,''
  Nucl.\ Phys.\ B {\bf 536}, 199 (1998)
  [arXiv:hep-th/9807080].
  %%CITATION = HEP-TH 9807080;%%

\bibitem{wl1}
  J.~M.~Maldacena,
  ``Wilson loops in large N field theories,''
  Phys.\ Rev.\ Lett.\  {\bf 80}, 4859 (1998)
  [arXiv:hep-th/9803002].
  %%CITATION = HEP-TH 9803002;%%

\bibitem{wl2}
  S.~J.~Rey and J.~T.~Yee,
  ``Macroscopic strings as heavy quarks in large N gauge theory and  anti-de
  Sitter supergravity,''
  Eur.\ Phys.\ J.\ C {\bf 22}, 379 (2001)
  [arXiv:hep-th/9803001].
  %%CITATION = HEP-TH 9803001;%%

\bibitem{wl3}
  S.~J.~Rey, S.~Theisen and J.~T.~Yee,
  ``Wilson-Polyakov loop at finite temperature in large N gauge theory and
  anti-de Sitter supergravity,''
  Nucl.\ Phys.\ B {\bf 527}, 171 (1998)
  [arXiv:hep-th/9803135].
  %%CITATION = HEP-TH 9803135;%%

\bibitem{wl4}
  A.~Brandhuber, N.~Itzhaki, J.~Sonnenschein and S.~Yankielowicz,
  ``Wilson loops in the large N limit at finite temperature,''
  Phys.\ Lett.\ B {\bf 434}, 36 (1998)
  [arXiv:hep-th/9803137].
  %%CITATION = HEP-TH 9803137;%%

\bibitem{wl5}
  J.~Sonnenschein,
  ``What does the string / gauge correspondence teach us about Wilson loops?,''
  arXiv:hep-th/0003032.
  %%CITATION = HEP-TH 0003032;%%


\bibitem{hl} H.~Liu, private communication. 


\end{thebibliography}
\end{document}